\documentstyle[12pt]{article}
\begin{document}
\begin{center}
{\bf Approximate integrals of motion and the quantum chaoticity problem\\
 V. E. Bunakov, I. B. Ivanov\\{\it Peterburg Nuclear Physics Institute}}

{\bf Abstract}
\end{center}
{\small
The problem of existence and constructing of integrals of motion in stationary
quantum mechanics and its connection with quantum chaoticity is discussed.
It is shown that the earlier suggested quantum chaoticity criterion
characterizes destruction of initial symmetry of regular system
and of basis quantum numbers under influence of perturbation.
The convergent procedure allowing to construct approximate integrals of motion
in the form of non-trivial combinations depending on operators $(q,p)$ is suggested.
Properties of the obtained integrals with complicated structure
and the consequences of their existence for system's dynamics are discussed.
The method is used for explicit construction and investigation of the
approximate integrals in Henon-Heiles problem.
}

\section{Introduction}

For last decades the investigations of quantum chaos have been extensively
carried out but this field remains under hot discussions. One part
of researchers believe that quantum chaos doesn't exist and at best should
be studied in the semiclassical approximation. It is obvious from correspondence
principle that quantum counterpart of classical system should have properties
reflecting regularity or chaoticity of classical trajectories. The law of
level spacing distribution is considered to be one of such properties.
It is believed that the quantum analogue of the chaotic classical obeys
Wigner level spacing law, while Poissonian law holds for regular systems.
However many authors (including us [1]), pointed out the
incompleteness and crudeness of this criterion of quantum chaoticity.

In this paper we continue to develop our approach [2-4] to the chaotic
properties of the quantum Hamiltonian systems. Our main point is connection
between the symmetry properties of a system and its regularity or chaoticity.
We show that the earlier suggested chaoticity criterion characterizes the
initial symmetry breaking and destruction of the corresponding integrals of
motion in a perturbed system, which leads to chaotisation. We compare our
approach with known criterion of existence  of the approximate
quantum numbers by Hose and Taylor [5, 6] which is based on the analysis
of effective Hamiltonians.

One may ask if the new integrals of motion might appear in the
perturbed system and how can they influence the system's dynamics.
The problem of construction of the approximate integrals of motion has
always attracted great attention due to its practical and
philosophical importance. In classical mechanics we have KAM-theory
which guarantees existence of invariant tori under perturbation;
as to quantum mechanics the situation is somewhat
tangled [5, 6]. The normal form method [7, 8] is well known for construction
of the new approximate integrals of
motion and integrable approximations to Hamiltonian.
Its generalization on quantum systems was done in refs. [9-11]. After
analyzing the reasons of divergence in the normal form method we propose a
rather simple way for construction of approximate quantum numbers based on the
unitary transformation of the basis integrals of motion.

In the final part of the paper we discuss the possibility of the exact
integrals
existence in the perturbed system and formulate the hypothesis
of formal integrability in stationary quantum mechanics. We define formal
integrability as the existence (in mathematical sense) of complete set of
independent mutually commuting operators which have some characteristic
differences from the usual first integrals of motion. It is impossible to
write such operators in closed form (they have extremely complicated
structure) and we can't find them before the solution of Schr\"odinger
equation is obtained;
these integrals don't correspond to known symmetry groups and are
non-separable. We discuss why such a formal (mathematical) integrability
 doesn't contradict the system's  quantum chaoticity. The conception
of chaos means that the system has algorithmic complexity (it is very
difficult to solve equations with high accuracy) and statistical hypothesis works well
in the system. The existence of non-separable integrals of motion with
very complicated structure can't help us to solve equations and doesn't
influence the validity of statistical hypothesis.

\section{Destruction of quantum numbers and the chaoticity criterion}

Presently we consider stationary quantum system with Hamiltonian $H$
as a sum of Hamiltonian $H_0$ of an integrable system and
perturbation $\lambda V$:
\begin{equation}
\label{H}
H=H_0+\lambda V.
\end{equation}
Eigenfunctions $\{\phi_\alpha\}$ of the unperturbed integrable
Hamiltonian $H_0$
are common for some complete set of independent mutually commuting
operators $\{J_\rho\}_{\rho=1}^N$ ($N$ - the number of degrees of
freedom of the Hamiltonian $H_0$). Eigenstates $\psi_i$ of the full
Hamiltonian may be expanded in eigenfunctions $\phi_\alpha$
of the unperturbed Hamiltonian $H_0$:
\begin{equation}
\label{expansion}
\psi_i=\sum_\alpha \phi^*_\alpha\psi_i\phi_\alpha =
        \sum_\alpha c^{\alpha}_i\phi_\alpha.
\end{equation}

Let us consider the probability $P_\alpha(E_i)$ to find the basis state
$\phi_\alpha$ in the state $\psi_i$ with energy $E_i$,
which is equal to the squared
absolute value of the corresponding coefficient in (\ref{expansion})
and define the energy width $\Gamma_{spr}^\alpha$ of $P_\alpha(E_i)$ distribution:
the minimal energy interval for which the sum of probabilities
$P_\alpha(E_i)$ is larger or equal to $0.5$.
Thus defined $\Gamma_{spr}^\alpha$ is the energy spreading width of basis
state $\phi_\alpha$. The spectrum of $H_0$ may be degenerate
and then the irreducible representations of symmetry group of $H_0$
consist of several basis functions which belong to one energy level (shell).
We want to find a parameter characterizing the measure
of initial symmetry breaking of $H_0$ under the influence of
the perturbation $V$. It's clear that
such symmetry breaking results only due to significant mixing between
functions from different irreducible representations. The mixing of
states within one shell doesn't change their symmetry. Unless the
spreading width $\Gamma_{spr}^\alpha$ is smaller then the distance $D_0$
between the neighboring levels in the spectrum of $H_0$ we can distinguish
"localization domain" (in energy) of one set of basis states from
"localization domain" of another one. When the spreading width exceeds
$D_0$ we start loosing the "signatures" of basis functions in the spectrum of
$H$ and can't even approximately compare states $\psi_i$ with irreducible
representation of symmetry group $H_0$. Thus parameter
\begin{equation}
\label{kappa}
\ae^\alpha = \Gamma_{spr}^\alpha /D_0
\end{equation}
is the natural measure of symmetry breaking.
When the parameter $\ae^\alpha$ exceeds unity the symmetry of the
Hamiltonian $H_0$ disappears. Such a value of perturbation is accompanied by
disappearance of the initial selection rules, the levels are distributed
approximately uniformly (level repulsion) and the level spacing
distribution approaches Wigner's law.
One can say that the transition from regularity to chaoticity
has taken place in quantum system and $\ae^\alpha$
may be considered as the parameter of chaoticity.

The spreading width $\Gamma_{spr}^\alpha$ depends on the number $\alpha$
of basis state of Hamiltonian $H_0$, i.e. on its quantum numbers.
The classical analogy to this is the dependence of the
invariant torus stability on the corresponding values of integrals of motion.
It is clear that in order to obtain the global chaoticity
characteristic in quantum
case it is necessary to average $\Gamma_{spr}^\alpha$ over the basis states
$\phi\alpha$ belonging to the same irreducible representation (shell).
The averaged chaoticity parameter $\Gamma_{spr}$
unlike local $\Gamma_{spr}^\alpha$
has an important feature of invariance with respect to the choice of the basis
for integrable Hamiltonian $H_0$ [12]. From the theoretical point of view the
above chaoticity parameter has one more useful property. As it was shown in
[3], in the semiclassical limit $\Gamma_{spr}^\alpha/\hbar$ transforms into
Lyapunov's exponent of the corresponding classical motion.

Thus we see that criterion $\Gamma_{spr}$
"measures" destruction (fragmentation) of the irreducible representations of the
basis and, hence of the Casimir operator (the main quantum number), which is
the approximate integral of motion for small perturbations. The other quantum
numbers of the basis states suffer destruction in general under a
smaller perturbation due to the strong mixing within one shell.
To define directly the degree of destruction of approximate integrals
of motion one may use an ordinary mean-square-root deviation from the mean
value. This deviation of some operator $A$ in the state
$\psi_\alpha$ is calculated as follows:
\begin{equation}
\label{dA}
\Delta A = \sqrt{\psi_\alpha^*(A - \psi_\alpha^*A\psi_\alpha)^2\psi_\alpha}.
\end{equation}
The operator $A$ is the approximate integral of motion if the ratio
of $\Delta A$ to
the difference between its neighboring eigenvalues is less then $1$.

Another way to study and construct the approximate integrals of motion is the
well known method of effective Hamiltonians. In the series of works Hose
and Taylor [5, 6] suggested the criterion of existence of the effective
Hamiltonian and of connected with it integrals of motion. According to this
criterion we may build the convergent sequence of approximations to the
effective
Hamiltonian under the condition that projection of perturbed Hamiltonian
wave function to the model
space is greater then 0.5. Thus if projection of some states $\psi_\alpha$
of Hamiltonian $H$ to the shell space exceeds 0.5, then the main quantum
number has to be the approximate integral of motion for these states. It
is obvious that this criterion practically coincides with our spreading
width one.

In order to compare the above criteria of destruction of the integrals of
motion we have analysed the quantum Henon-Heiles system:
\begin{equation}
\label{hhh}
H(q,p) = \frac{1}{2}(p^2_1+q^2_1)+\frac{1}{2}(p^2_2+q^2_2) +
\lambda (q^{2}_1 q_2 - q^3_2/3)
\end{equation}
The eigenfunctions were obtained using the oscillator basis of $496$
states ($30$ shells). Fig.1 shows the dependence on the perturbation
intensity (energy $E$) of the parameter $\kappa$, the averaged spreading
width of operator $N=n_x+n_y$ (see \ref{dA}) and of the
averaged projection ($P_s$) of the exact wave functions to the shell.
For the sake of easier comparison with other quantities in Fig.1 we
plotted the value $Pr=2(1-P_s)$.
We see that destruction of the initial $SU(2)$-symmetry according
to all these three criteria takes place approximately at $E = 0.10$.

Thus $\Gamma_{spr}$ measures the degree of mixing between the irreducible
representations and the destruction of the corresponding Cazimir operators.
The question arises if some new integrals of motion might appear instead of
the destroyed basis integrals.

The method of normal forms is the most known way to construct the
approximate integrals and the integrable
approximations to Hamiltonian in classical mechanics [7, 8]. It has been
generalized
on quantum systems in refs. [9-11]. In this method the perturbed wave
functions
are constructed as certain superposition of the basis functions belonging
to a single irreducible representation. Therefore the symmetry of the wave
functions is not changed and perturbation $V$ with lower symmetry
leads only to the splitting of the degenerated level.

This approach gives rather good results when the
perturbation is small.
However Siegel (1941) proved the divergence of classical
normal forms in the case of nonintegrable initial system.
In quantum mechanics the question about
convergence hasn't got a final solution, though the authors [10, 11]
stress the asymptotic character of the series arising
and confirm it by numerical calculations.

Two reasons for the divergence in the method of normal forms may be
pointed out. The first one is well known --- it is the non-analyticity
of solutions at the point $\lambda=0$
(the replacement $\lambda \to -\lambda$ in Hamiltonian may lead to significant
changes of spectrum properties) and, as a consequence, to the divergence of
the expansion
into the powers of $\lambda$. The second reason is as follows.
In the quantum Birkgoff-Gustavson method the
integrable approximations are constructed in the
form of power series of operators which mix basis states only inside one
irreducible representation of the basis symmetry group (one shell).
If the interaction effectively mixes states from different irreducible
representations the Birkgoff-Gustavson
method obviously can't generate good integrable
approximations and convergent integrals in principle. This is
why the quantum numbers given by normal form might be good only when mixing
between the different shells is weak ($\Gamma_{spr}<1$).

The difficulties described doesn't mean the principal absence of
approximate integrals of motion in the perturbed system, they only reveal the
shortcomings of the methods. To get a convergent method of constructing the
integrals one should use combinations of
operators which mix states in any given finite-dimensional subspace.
In this case we can improve the integrable approximation to $H$
(in the sense of operator norm) by simply increasing the dimensions of
this subspace.
In the next section we'll describe the convergent procedure allowing
to construct approximate integrals in a rather trivial way.

\section{The convergent method for integrals of motion}

Performing a unitary transformation $U$ of some basis with a set of quantum
numbers we can always easily find a new set of mutually commuting
operators for which a transformed basis functions are eigenfunctions.
One can see that $J'_\rho = UJ_\rho U^\dagger$
($J_\rho$ -- operators of initial basis) are the desirable operators
and such bases are equivalent in the sense of the quantity of quantum numbers.
If we assume the completeness of eigenstates $\psi_\alpha$ of Hamiltonian $H$
in the Hilbert space $\cal H$ then $\psi_\alpha=U\phi_\alpha$
($\phi_\alpha$ is the eigenfunction of $H_0$), because any two complete
orthonormal bases are connected via some unitary transformation $U$.
Operators $J'_\rho$ commute with $H$ in the full Hilbert space $\cal H$.
Actually, for any function from complete basis $\psi_\alpha$
$$
[H, J'_\rho]\psi_\alpha = HUJ_\rho U^\dagger\psi_\alpha -
UJ_\rho U^\dagger H\psi_\alpha = HUJ_\rho\phi_\alpha -
E_\alpha UJ_\rho\phi_\alpha =
$$
$$
j_{\rho\alpha}HU\phi_\alpha -
E_\alpha j_{\rho\alpha} U\phi_\alpha =
j_{\rho\alpha}E_\alpha\psi_\alpha - E_\alpha j_{\rho\alpha}\psi_\alpha = 0.
$$
$J'_\rho$ are derived from $J_\rho$ with the aid of unitary transformation,
hence they also form a complete set of independent commuting operators
and their eigenvalues $j_{\rho\alpha}$ uniquely define every eigenstate
$\psi_\alpha$ of Hamiltonian $H$. Therefore $H$ is a function depending
on operators $J'_\rho$: $H=H(J'_\rho )$, and $E_\alpha = H(j_{\rho\alpha})$.
Having determined the approximate wave functions of perturbed Hamiltonian we
can construct the unitary transformation $U$ and the
approximate integrals of motion $J'_\rho$. The question about convergence
of the procedure suggested is reduced to investigating whether the
corresponding method for solving of Schr\"odinger equation converges or not.
For example it has been proved that Ritz's method converges
(with increasing basis) in the case of Hermitian operators
with lower-bounded discrete spectrum [13].

The introduced operators $J'_\rho$
seem to be formal unless we construct them explicitly as functions
of dynamic variables $(q,p)$. This, however is not difficult to do
with the help of methods taken from the theory of continuous groups'
representations. In the remaining part of this section the realization
of this method is described in details and properties of the integrals
obtained are discussed.

Let us consider Schr\"odinger equation $H\psi=E\psi$ with discrete spectrum
$E_\alpha$ and $\psi_\alpha$ and write the Hamiltonian $H$ in the form of
spectral decomposition
\begin{equation}
\label{Hexp}
H=\sum_\alpha E_\alpha\psi_\alpha\psi_\alpha^*.
\end{equation}
We represent the complete Hilbert space $\cal H$ as a sum of the
finite-dimensional model space $\cal P$ and its orthogonal
adjunct $\cal Q$:
$\cal {H = P + Q}$. It is convenient to construct the space $\cal P$
of eigenfunctions $\{\phi_\mu\}_{\mu=1}^{dim\cal P}$ by using
some complete set of mutually commuting operators $\{J_\rho\}_{\rho=1}^N$
($N$ - the number of degrees of freedom $H$). We'll find the
approximate wave functions of $H$ in the $\cal P$-space
as a combination of basis states $\phi_\mu$ (as, for example, in Ritz's
variational method or in different versions of perturbation theory)
\begin{equation}
\label{stexp}
\psi_{p \alpha}=\sum_{\mu\in\cal P}c_\alpha^\mu\phi_\mu.
\end{equation}
Orthonormal states $\psi_{p \alpha}$ are derived from minimum condition for
the energy
functional in the $\cal P$-space and they form a
subspace in the $\cal P$-space (obviously,
only a small number of combinations (\ref{stexp}) will satisfy Schr\"odinger
equation with sufficient accuracy). We'll denote this subspace of solutions
by $\cal S$ and the energy of states by $E_{p \alpha}$.
The rest $dim{\cal P} - dim{\cal S}$ of the basis functions in the
$\cal P$-space may be chosen arbitrarily, and we denote the new basis in
the $\cal P$-space $\{\phi'_\mu\}_{\mu =1}^{dim\cal P}
(\phi'_\mu = \psi_{p\mu}, \mu =1,...,dim\cal S)$.
Let us show that the operator
\begin{equation}
\label{Hs}
H_s=\sum_{\alpha\in\cal S} E_{p\alpha}\psi_{p \alpha}\psi_{p \alpha}^*
\end{equation}
(i) commutes with operators forming a complete set in the full Hilbert
space $\cal H$, i.e. it is integrable, (ii) $H_s$ is approximating
$H$ in the sense of operator norm in the $\cal S$-space,
(iii) $H_s$ may be expressed in terms of dynamic variables $(q,p)$ as well
as the initial Hamiltonian $H$.

By calculation of wave functions $\psi_{p \alpha}$
we constructed simultaneously the unitary transformation
$\phi'_\mu = U\phi_\mu$ of space $\cal H$, which is defined
by coefficients $c_\alpha^\mu$ in the $\cal P$-space and
is the identical transformation in the $\cal Q$-space. Operators
$J'_\rho = UJ_\rho U^\dagger$ are known to form the complete set with
the same quantum numbers $j_{\rho\alpha}$ and eigenfunctions $\phi'_\mu$.
As far as eigenfunctions of operators $J'_\rho$ and $H_s$ in the
$\cal S$-space coincide, the operators commute in this space. Outside the
$\cal S$-space  $H_s\equiv 0$ and hence it also commute with $J'_\rho$.
Therefore $[H_s, J'_\rho]=0$ in the full space $\cal H$.

Now we are going to check that Hamiltonian $H_s$ is close to $H$ in the sense
of the operator norm in $\cal H$, i.¥.
$||H-H_s||_{\cal S} < \epsilon$, under the condition that the residual
of the approximate solutions
$(H-E_{p\alpha})\psi_{p\alpha} = \delta\psi_\alpha$
doesn't exceed $\epsilon$: $||\delta\psi_\alpha|| < \epsilon$.
Really, for an arbitrary function  $\chi = \sum a^\delta\psi_{p\delta},\;
||\chi||=1,\;\chi\in\cal S$
$$
||H\chi - H_s\chi|| = ||(\sum_{\alpha\in\cal H}
E_\alpha\psi_\alpha\psi_\alpha^* - \sum_{\alpha\in\cal S}
E_{p\alpha}\psi_{p\alpha}\psi_{p\alpha}^*)\sum_{\delta\in\cal S}a^\delta
\psi_{p\delta}||=
$$
$$
||\sum_{\delta\in\cal S}a^\delta (\sum_{\alpha\in\cal H}
E_\alpha\psi_\alpha\psi_\alpha^*\psi_{p\delta} - E_{p\delta}\psi_{p\delta})||
\le\epsilon\sum_{\delta\in\cal S}|a^\delta|\le\epsilon\sqrt{dim\cal S}.
$$
In the last estimate we used the fact that $\sum|a^\delta|$
under the condition $\sum|a^\delta|^2 = 1$ reaches its maximum value when
all $a_\delta$ are identical and equal to $1/\sqrt{dim\cal S}$.
The accuracy $\epsilon$ depends on dimensionality of $\cal P$-space;
we may fix $dim\cal S$ and decrease $\epsilon$ in such way that the
norm $||H-H_s||_{\cal S}$ should be as small as we need.

The introduced operators $H_s$, $U$ and $J'_\rho$ seem to be formal unless we
construct them explicitly as functions of dynamic variables $(q,p)$.
Writing operator $\psi_{p \alpha}\psi_{p \alpha}^*$ in terms
of expansion (\ref{stexp}) we have:
\begin{equation}
\label{Hsb}
H_s=\sum_{\mu,\nu\in\cal P}\Bigl\{\sum_{\alpha\in\cal S}
E_{p\alpha}c_\alpha^\mu {c_\alpha^{\nu}}^*\Bigl\}\phi_\mu\phi_{\nu}^*.
\end{equation}
The operator $U$ may be also represented as a combination of basis
operators $\phi_\mu\phi_{\nu}^*$:
\begin{equation}
\label{U}
U= 1+\sum_{\alpha\ne\beta\in\cal P}U_{\alpha\beta}
\phi_{\beta}\phi_{\alpha}^{*}+\sum_{\gamma\in\cal P}
(U_{\gamma\gamma}-1)\phi_{\gamma}\phi_{\gamma}^{*}.
\end{equation}
The first $dim\cal S$ rows of the unitary matrix $U_{\alpha\beta}$
coincide with matrix $c_\alpha^\beta$, the rest
$dim{\cal P} - dim{\cal S}$ may be chosen in arbitrary way. It is easy
to verify that components $\chi$ don't change outside $\cal P$ when
$U$ acts on arbitrary state $\chi=\sum a^\delta\phi_\delta$, while inside
$\cal P$ they are transformed by unitary matrix:
$$
U\chi= \sum_{\delta\notin\cal P}a^\delta\phi_\delta +
\sum_{\alpha\in\cal P}\Bigl(\sum_{\beta\in\cal P}
U_{\beta\alpha}a^\beta\Bigr)\phi_\alpha.
$$

Now we'll construct the basis operators $\phi_\mu\phi_{\nu}^*$ as ordinary
operators in the form of combinations depending on variables $(q,p)$
and acting on states of Hilbert space in co-ordinate representation.
If $\cal G$ is the group of transformations corresponding
to the complete operator set $J_\rho(q,p)$, then $\cal P$-space
is in general a direct sum of irreducible representations ${\cal T}_s$ of
the group $\cal G$:
$$
{\cal P} = \sum_s\oplus {\cal T}_s.
$$
Operator $\phi_\mu\phi_{\nu}^*$ transforms the function $\phi_\nu$
into $\phi_\mu$. If the group $\cal G$ is Abelian one, the irreducible
representations ${\cal T}_s$ are one-dimensional and consist of the
function $\phi_s$. For non-Abelian group $\phi_\mu$ and $\phi_\nu$ may
belong to one irreducible representation.
Our aim is to write the operators which generate all possible
transitions between different ${\cal T}_s$ and within some ${\cal T}_s$
as well. It's not difficult to solve this problem by methods of group theory
and actually the problem is equivalent to the realization of the basis
$\phi_\mu$.

To perform transformations inside the irreducible representations
${\cal T}_s$ it's sufficient to use combinations of generators of the basis
symmetry group; to connect different ${\cal T}_s$ we need generators
of special non-invariance group of basis. Its infinite-dimensional
irreducible representation is spanned on our basis. We know
non-invariance groups and corresponding algebras for various bases, for
example,
$so(4,2)$ --- for Coulomb's basis and $so(3,2)$--- for isotropic
two-dimensional oscillator basis [14].
After realization of operators of non-invariance algebra $A_\alpha$ in the
form of combinations of dynamic variables $(q,p)$ we look for vacuum
state $\phi_0$ for which the decreasing operators from the set
$A_\alpha (q,p)$ give zero. The vacuum state forms one-dimensional
irreducible representation of symmetry group $\cal G$, and we shall
naturally obtain states from other irreducible representations
$\cal G$ acting on it with creation operators from the
set $A_\alpha$. Notation $S^\dagger _\mu(A_\alpha)$ defines the
operator composed of generators $A_\alpha$ which produces
basis function $\phi_\mu$:
$\phi_\mu = S^\dagger _\mu \phi_0$ ¨ $S_\mu\phi_\mu = \phi_0$.
We don't present the general formula for $S^\dagger _\mu =S^\dagger _\mu
(A_\alpha)$ because it is not difficult to do it in any specific case
(see section 4); usually $S^\dagger _\mu$ are
polynomials composed of generators
$A_\alpha$ the power of which increases with state's number $\mu$.
Then operator $\phi_\mu\phi_{\nu}^*$ on the
Hilbert space $\cal H$ may be written as follows:
\begin{equation}
\label{baop}
\phi_\mu\phi_{\nu}^* = S^\dagger _\mu S_\nu P_\nu,
\end{equation}
where $P_\nu$ is the projector on the state $\phi_\nu$.
The projector $P_\nu$ may also be expressed in terms of dynamic variables
$(q,p)$ in the following way. Let $T(x)$ be the operators of unitary
representation $\cal G$ in Hilbert space,
$D^{s}_{\alpha\beta}(x)$ --- matrix elements of
irreducible representation ${\cal T}_s$, $dx$ --- invariant Haar's measure
on $G$. Then the projector $P_\nu$ on the basis state $\phi_\nu \in
{\cal T}_s$ may be presented as [15]:
\begin{equation}
\label{proekt}
P_\nu = dim{\cal T}_s\int_G dx{D^{s *}_{\nu\nu}}(x)T(x).
\end{equation}
Operators (\ref{proekt}) are bounded, and since $T(x)$ is the exponent to the
power of generators $\cal G$ which form sub-algebra with respect to
operator algebra $A_\alpha (q,p)$, we have achieved our goal ---
expressed the basis operators $\phi_\mu\phi_{\nu}^*$,
$H_s$ (\ref{Hsb}), $U$ (\ref{U}) and integrals $J'_\rho$,
in terms of variables $(q,p)$.

Thus constructed integrable approximations $H_s(q,p)$ and
integrals of motion $J'_\rho(q,p)$, apart from their approximate character
(the commutators with $H$ are not exactly equal to zero) are local.
The Hamiltonian $H_s$ is close to $H$ in the sense
of operator norm only in the finite-dimensional
subspace $\cal S$, while the operators $J'_\rho$ are good
invariants also only in the $\cal S$-space ( outside the
$\cal P$-space they coincide with the old operators $J_\rho$).
In the following section we shall demonstrate how the method works in
Henon-Heiles problem and then continue to discuss the
properties of the integrals obtained.

\section{Approximate integrals in Henon-Heiles problem}

Here we apply the method of the integral construction developed in the
previous
sections to the well known Henon-Heiles problem with Hamiltonian (\ref{hhh}).
Introducing operators of creation and annihilation
$a^\dagger _k = \frac{1}{\sqrt{2}}(q_k + ip_k),
a_k = \frac{1}{\sqrt{2}}(q_k - ip_k), k = 1,2$ we construct, as usual,
the Cartesian oscillator basis
\begin{equation}
\label{cob}
\phi_\mu = \phi_{n_1n_2} = \frac{1}{\sqrt{n_1!n_2!}}
(a^\dagger _1)^{n_1}(a^\dagger _2)^{n_2}\phi_0
\end{equation}
We present the projector
$P_{n_1n_2}$ on the state $\phi_{n_1n_2}(q_1, q_2)
=\phi_{n_1}(q_1)\otimes\phi_{n_2}(q_2)$
as a product of projectors
on the states $\phi_{n_1}(q_1)$ and $\phi_{n_2}(q_2)$ of the corresponding
one-dimensional oscillator. For one-dimensional oscillator
${\cal G} = U(1),\;T(x) = e^{ia^\dagger ax}, \;
x \in [0, 2\pi ], \; D^s_{\nu\nu} = e^{i\nu x}$, where
the number of state $\nu$ equals the number of quanta in this state.
Then according to (\ref{proekt})
$$
P_n = \int_0^{2\pi}\frac{dx}{2\pi}e^{-inx}e^{ia^\dagger ax}=
-\frac{i}{2\pi}(a^\dagger a - n)^{-1}\{e^{2\pi i(a^\dagger a - n)} - 1\}=
$$
$$
\frac{1}{\pi}(a^\dagger a - n)^{-1}e^{i\pi (a^\dagger a - n)}
\sin{\pi (a^\dagger a - n)}=
\frac{1}{\pi}(a^\dagger a - n)^{-1}\sin{\pi (a^\dagger a - n)}.
$$
We neglect the phase in the last expression because it does not affect the
action of $P_n$. It's easy to check that $P_n$
acts in the necessary way due to its slightly exotic form:
$$
P_n\phi_\mu = \delta _{n\mu}\phi_\mu.
$$
The total projector $P_{n_1n_2}$ takes the form
\begin{equation}
\label{SU(2)prkt}
P_{n_1n_2} = \frac{1}{\pi^2}(a^\dagger _1 a_1 - n_1)^{-1}
(a^\dagger _2 a_2 - n_2)^{-1}\sin{\pi (a^\dagger _1a_1 - n_1)}
\sin{\pi (a^\dagger _2a_2 - n_2)}.
\end{equation}
As a result the formulae (\ref{baop}) together with (\ref{cob}) gives us the
operator $\phi_\mu\phi_{\nu}^*$:
\begin{equation}
\label{hhbo}
\phi_\mu\phi_{\nu}^* =
\frac{(a^\dagger _1)^{n_1(\mu )}(a^\dagger _2)^{n_2(\mu
)}}
{\sqrt{n_1(\mu )!n_2(\mu )!}}\frac{(a_1)^{n_1(\nu )}(a_2)^{n_2(\nu )}}
{\sqrt{n_1(\nu )!n_2(\nu )!}}P_{n_1(\nu )n_2(\nu )},
\end{equation}
where $n_1$ and $n_2$ are the quantum numbers of states. Using (\ref{Hs}),
(\ref{U}) and determining coefficients $c^\mu _\alpha$ and $E_{p\alpha}$ we
get integrable approximation $H_s$ and approximate integrals
$J'_\rho = UJ_\rho U^\dagger$. We may take $n_1$ and $n_2$ or $n_1$
and $n = n_1 + n_2$ as independent integrals $J_\rho (\rho = 1,2)$.
The coefficients of expansion $c^\mu _\alpha$ and the
energies $E_{p\alpha}$ were calculated by Ritz's method with the aid of
diagonalization of matrix $H$ on the basis (\ref{cob}).
We shan't write explicit expressions for $J'_\rho$ for they are
very cumbersome: if $D$ is the basis dimensionality,
then operators $J'_\rho$ consist of $D^4$ terms of the type
(\ref{hhbo}). To determine the degree of destruction of
approximate integrals of motion $J'_\rho$ in the states $\psi_\alpha$ we
calculate the mean-square-root deviation $J'_\rho$ with the help of
(\ref{dA}). The solutions obtained with the basis
of $496$ states ($30$ shells) were considered to be true wave functions.
Fig.2 shows the dependence of averaged measure of destruction
(\ref{dA}) of operators $J'_\rho$
($J_1 = l$ ¨ $J_2 = n$) on the perturbation intensity (energy $E$) in the
subspaces with different symmetries. Henon-Heiles Hamiltonian has a symmetry
$C_{3v}$. Therefore the eigenfunctions' space can be divided into 4 subspaces
$A, B, C, D$ (the states belonging to $C$ and $D$ subspaces have the same
energy and thus produce the sequences of degenerate levels). The approximate
integrals $J'_\rho$ were calculated for the $\cal P$-space of different
dimensionalities (1, 10, 15 and 20 shells).
 One can see that
increasing of $\cal P$-space dimensionality is accompanied
by the decrease of fragmentation of the approximate integrals in the
$\cal S$-space; outside the $\cal S$-space operators
$J'_\rho$ loose their advantages in comparison with $J_\rho$.

This example shows that the approximate integrals $J'_\rho$
really have smaller spreading (fragmentation) in the $\cal S$-space
then the basis integrals of motion $J_\rho$. We may get very small
values $\Delta J'_\rho\approx 0$ for the bounded states
by increasing of $dim\cal P$. As a result the analytical structure of
$J'_\rho$ becomes very complicated.

\section{Are integrability and chaos compatible?}

Now we consider the question about the convergence of the suggested
procedure and the question about existence of exact integrals of motion
of Hamiltonian $H$ in the full Hilbert space (integrability of $H$).
With the help of Ritz's method we can find in principle any finite number of
states with any finite accuracy and thus construct integrable approximation
$H_s$ to $H$ and integrals $J'_\rho$ in any finite subspace with any
desirable accuracy. The question is: can we obtain the full infinite
spectrum of $H$ by tending $\cal P \to \cal H$, because
there is an effect of systematic "delay" of $\cal S$-space
dimensionality with respect to dimensionality of
the model $\cal P$-space (only states far from the boundary of the
approximate spectrum are reasonably accurate in diagonalization).
In other words, does the sequence of Hamiltonians $PHP$ converge to the
initial
Hamiltonian: $PHP \to H$ while $P\to 1$. Physically it seems to be so
but we are in a difficulty to give rigorous mathematical proof,
because the sequence $PHP$ is not Caushy's sequence in the sense of
operator norm in $\cal H$.

We can formulate the following hypothesis about integrability of $H$.
If there exist a good (in mathematical sense)
unitary operator $U$ connecting two complete
orthonormal bases: the initial $\phi_\alpha$ and the basis of eigenstates
$\psi_\alpha$, then the Hamiltonian $H$, as we have seen above, commutes with
the complete set of independent operators $J'_\rho$ and it may be expressed
in terms of only these variables $H=H(J'_\rho )$. (Moreover, according to
Dirac [16] any functions of arbitrary complete orthonormal basis are
eigenfunctons of some complete set of commuting observables. This allows to
extend our conclusions for any observable, including the case of continuum
spectra). Therefore the Hamiltonian $H$ seems to be formally integrable.
However, in the above considered
example (the Henon-Heiles Hamiltonian) the system is chaotic according
to all the criteria of Section 2. Moreover, we know for sure that in the
classical limit this system is one of the textbook examples of chaoticity.
The problem is how to remove the contradiction between the seeming formal
integrability and the chaoticity of the system.

One possible answer is connected with the properties of the new integrals
$J'_\rho$.
These integrals of motion
are independent and global (provided the convergence of $PHP$ discussed above
would be proved). However they have extremely complicated structure
and can't be expressed in closed form. Therefore they are useless
to separate variables (non-separable) and to solve a problem.
We restore them after the approximate numerical solution has been found.
These integrals don't give selection rules for transitions between levels.
Therefore they are definitely not the quantum analogs of the classical first
(isolating) integrals, which define the classical regular integrable system.

\section{Conclusions}

The problem of existing and constructing of integrals of motion in
stationary quantum mechanics and its connection with notion of
quantum chaoticity has been investigated. It has been shown that the
previously suggested quantum chaoticity criterion characterises
destruction of initial symmetry of regular system and basis
integrals of motion under the influence of perturbation. Our approach
conforms with known probability criterion of Hose and Taylor [5-6]
and direct estimate of fragmentation (\ref{dA}).

We use variational Ritz's method for explicit construction
of approximate integrals of motion in the form of combinations
depending on operators $(q,p)$ though in principle
another method for solving Schr\"odinger equation may be used.
As a result we obtained finite large-dimensional sums consisting of
non-invariance algebra operators in various powers and projectors
nontrivially expressed in terms of invariance algebra generators.
The quality of approximate integrals of motion is simply controlled
by dimensionality of model space in use.

These integrals of motion are independent and global (provided the
convergence of $PHP\to H$ discussed above would be proved).
However they have extremely complicated structure but
have extremely complicated structure
and can't be expressed in closed form, therefore they are useless
to separate variables (nonseparable). This also explains why the
existence of these integrals doesn't create obstacles to statistical
description of quantum system. That's why such formal integrability
(even if we'll prove the existence of global integrals rigorously)
doesn't make system to be regular in the sense of absence of chaotical
properties. Therefore they are definitely not the quantum
analogs of the classical first
(isolating) integrals, which define the classical regular integrable system.

One of the authors (IBI) is indebted to Prof.Zikiki and to the Organizing
Committee of V.Gribov's Foundation for their support.

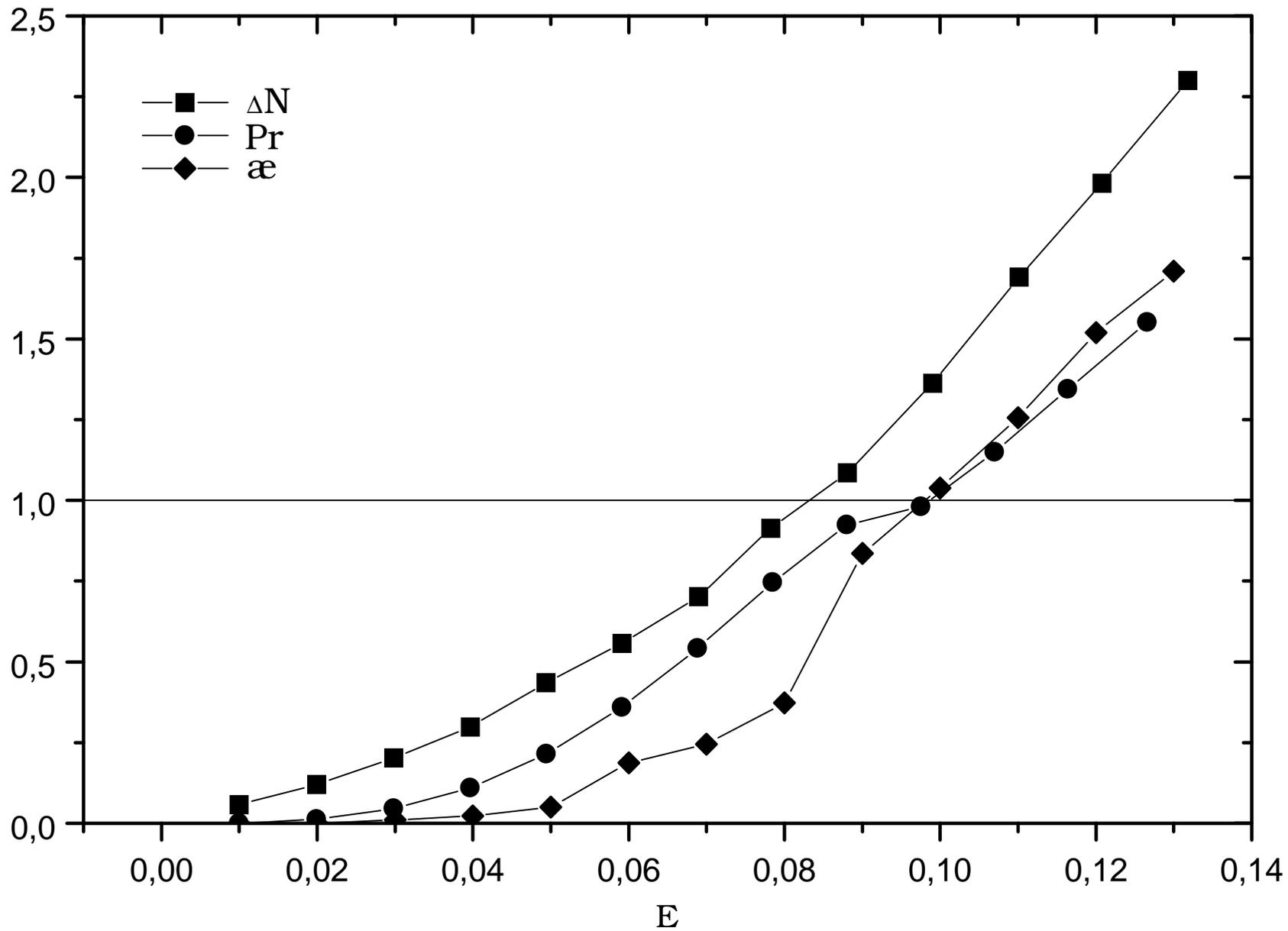
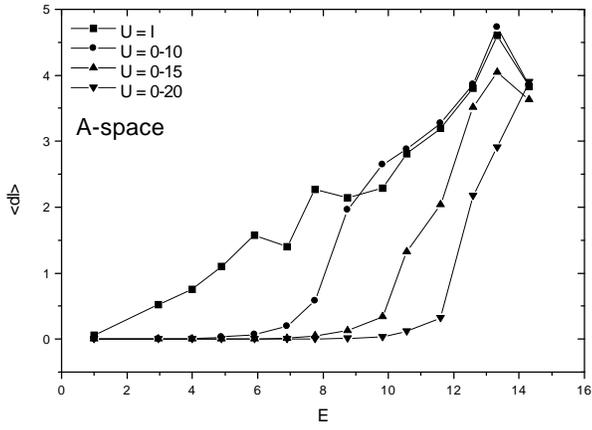
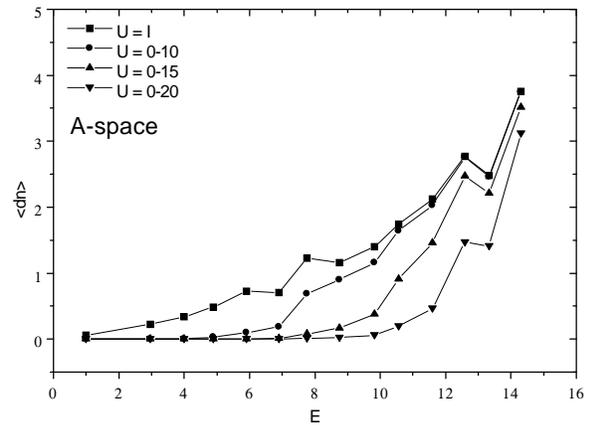
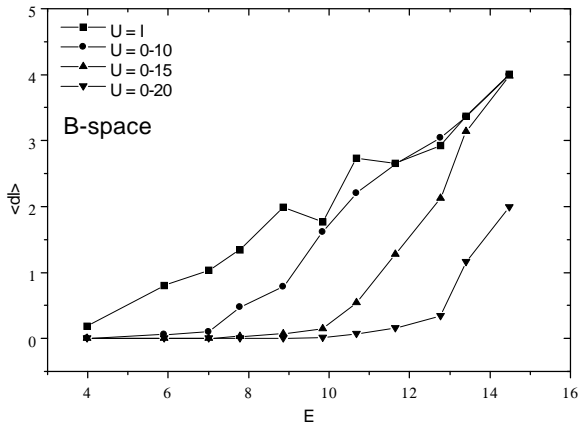
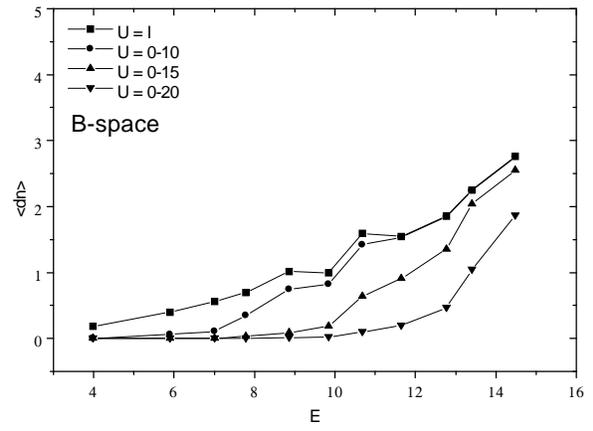
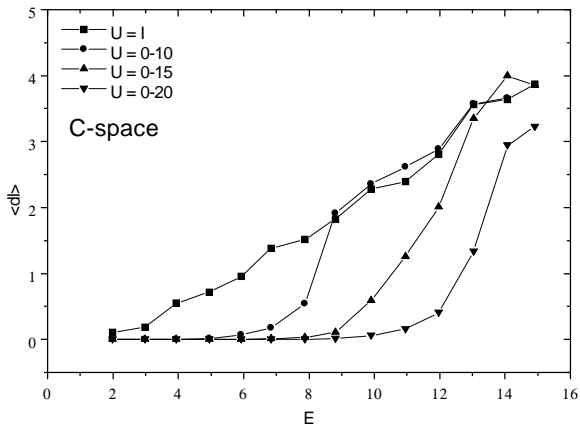
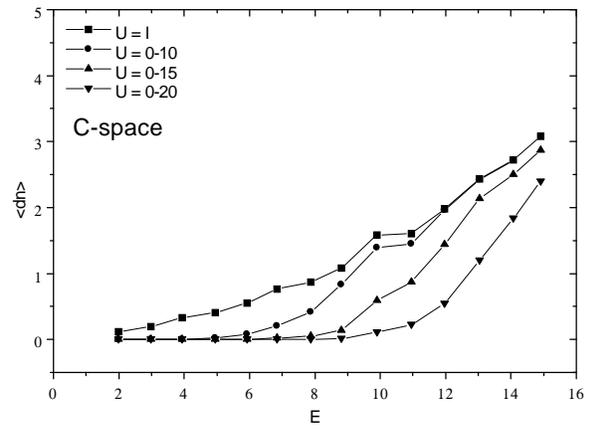
\end{document}